# Spreadsheets in Clinical Medicine - A Public Health Warning

Grenville J. Croll
Grenville@spreadsheetrisks.com
Director, Spreadsheet Engineering Ltd

Raymond J Butler
Ray.butler@virgin.net
EuSpRIG – European Spreadsheet Risks Interest Group

**ABSTRACT**

*There is overwhelming evidence that the continued and widespread use of untested spreadsheets in business gives rise to regular, significant and unexpected financial losses. Whilst this is worrying, it is perhaps a relatively minor concern compared with the risks arising from the use of poorly constructed and/or untested spreadsheets in medicine, a practice that is already occurring. This article is intended as a warning that the use of poorly constructed and/or untested spreadsheets in clinical medicine cannot be tolerated. It supports this warning by reporting on potentially serious weaknesses found while testing a limited number of publicly available clinical spreadsheets.*

## 1 INTRODUCTION

The impact and to some extent the incidence of errors in financial spreadsheet models have been widely documented [Panko, 2000, 2006] [Butler, 2000] [Croll, 2005] [EuSpRIG, 2006]. We are aware of no similar studies in medical domains.

Because spreadsheet users are human, spreadsheets are error prone. This has been shown by repeated studies of various types over several decades. Experiment has shown time and again that the only method for reducing errors in spreadsheets is the use of multiple people to test a spreadsheet, with multiple test passes. Even then, errors will remain.

Recent quantitative evidence shows that spreadsheet users are also overconfident [Panko, 2003]. Since they rarely test their spreadsheets, they don't find any errors, increasing their confidence in the way they use spreadsheets. If they do find an error or two, then that also increases their confidence, as they are not motivated to find further or all errors.

Finally, there is some evidence [Banks A., Monday D., 2002] that the differing ways that spreadsheet users interpret the real world and model it in a spreadsheet gives rise to differing numerical evaluations of the same situation.

The use of software in medicine is not new of course [Johnson, K.A, Svirbely, J.R., Sriram, et al], (2002)], what is new we believe are the issues related to the widespread use of end-user software, specifically spreadsheets, in medicine. We believe these issues to be of crucial importance when considering the continued use of spreadsheets to support potentially life or death decisions relating to patient care.

Medical error accounts for an estimated 98,000 deaths annually in the USA, 30,000 in the UK, and is the seventh largest cause of death [Kohn, L.T., Corrigan, J. & Donaldson, M. (Ed) (2000)].

## 2 SPREADSHEETS IN CLINICAL MEDICINE

A search of the PubMed abstracts database [PubMed, 2006] revealed over eight hundred references to the use of spreadsheets, many of which were clearly used in clinical applications [Maceneaney PM., Malone D.E., 2000] [Linthout N., et al, 2004] [Cederbaum M., Kuten A. 1999]. An internet (Google) survey of the word "spreadsheet" followed by a medical keyword such as "cardiovascular", "pediatrics", "anaesthesiology", "oncology" etc quickly identified a number of circumstances where publicly available spreadsheets are being used or could be used in clinical situations.

We assume that the spreadsheets identified in this manner represent a small fraction of the spreadsheets actually in use in medicine. Some evidence for this is found in reports [Johnson et al, 2002] that medal.org, a web site containing some thousands of downloadable medical spreadsheets, received in excess of 1,800 unique visitors per day in 2002.

### 2.1 Paediatrics

A recent article [Narchi, H. 2004a] is abstracted as follows:

> *"In a series of three articles, we describe the step-by-step design and use of a spreadsheet to analyze the results of a diagnostic test or a therapy in the literature. This first article describes the required skills, which are minimal. The hardware and software requirements are modest, widely available and relatively cheap. In addition to the elimination of the potential risk of calculation errors, time and effort is saved by the physician.* ***The use of such a spreadsheet will further consolidate the concept of evidence-based medicine by readers of the medical literature and will help to further improve the quality of care****"* (Authors' emphasis)

The last sentence in this abstract indicates that the spreadsheets referred to are to be intended for use in patient care. That the spreadsheets are intended for modification is made clear in the article by:

> *"The flexibility of the design also allows customization by each user, such as …..adding other statistical formulas for further analysis of data…."*

Modification of spreadsheets is a well known source of error. This arises through a variety of mechanisms, including the unintentional overwriting of formulas and the accidental entry of incorrect data and formulas.

A second article [Narchi, H., 2004b] describes a spreadsheet which implements Bayes theorem to compute the post-test probability of diagnosing a disease based on the prevalence of that disease in a clinician's practice. Table 1 in the article lists 23 cell addresses and their contents, being descriptive labels and a further 20 cell addresses and their contents, being formulas.

There are material risks of

- typographic error in such lists of formulas,
- error in the entry of the formulas into an actual spreadsheet.

The spreadsheet is of a similar size but substantially greater complexity than those used in studies in other domains used to determine the likelihood of making a mistake [Panko, R., 2003].

The following sentences give rise to concern:

> "Table 1 describes the data to be entered in particular cells.....Save your work. The spreadsheet is now ready for use"

Critical paragraphs about how the spreadsheet should be checked, tested, brought into use and maintained over its lifetime and through the many modifications it will endure are missing from this documentation.

**2.2 Anaesthesia**

A web site designed to inform Nurse Anesthetists [Evans, T.J., 2006] contains the following text:

> *".... here are two guides to help you in your anesthesia practice. First is a Microsoft Excel spreadsheet titled 'Pediatric Anesthesia Worksheet'. Use it to calculate medications and other parameters for pediatric patients".*

The spreadsheet pediatriccalV2.xls referred to above contains a data entry box where the user can enter their paediatric patient's Age, Weight, Height, Hours NPO, Respiratory Rate, Hematocrit and Minimum Allowable Hematocrit. The spreadsheet then calculates using a series of spreadsheet logic cells the doses of drugs for pre-operative, peri-operative and post-operative care including narcotics, analgesics, antibiotics, muscle relaxants and emergency medications.

The spreadsheet is protected, that is a password is ostensibly required to modify it. However, by simply cutting and pasting the whole spreadsheet to another worksheet, the spreadsheet is fully accessible in its entirety and therefore potentially at risk of untested modification. The part of the spreadsheet that calculates pre-operative doses is as follows:

|    | I | J | K | L |
|---|---|---|---|---|
|    |   |   |   | mg |
| 5  | *Pre-operative* |   |   |   |
| 6  | Atropine | IM | 0.02 mg/kg | 0.10 |
| 7  | Atropine | IV | 0.01 mg/kg | 0.10 |
| 8  | Cimetidine | PO/Slow IV | 7.5 mg/kg | 0.0 |
| 9  | Clonidine | PO | 4 mcg/kg | 0.000 |
| 10 | Glycopyrrolate | IV/IM | 0.01 mg/kg | 0.00 |
| 11 | Ketamine Stun | IM | 5 mg/kg | 0.0 |
| 12 | Metoclopramide | IV | 0.1 mg/kg | 0.0 |
| 13 | Midazolam | IV | 0.05 mg/kg | 0.00 |
| 14 | Midazolam | PO | 0.5 mg/kg | 0.0 |
| 15 | Midazolam | IM | 0.08 mg/kg | 0.00 |
| 16 | Midazolam | Nasal | 0.3 mg/kg | 0.0 |
| 17 | Morphine | IM | 0.1 mg/kg | 0.0 |
| 18 | Ranitidine | IV *(up to 50 mg)* | 1 mg/kg | 0.0 |

The formula for calculating e.g. atropine dose (L7) bears examination:

=IF(E19*0.02>0.6,0.6,IF(E19*0.02<0.1,0.1,E19*0.02))

E19 contains the weight in Kilograms. Perhaps if E19 had been defined as a name e.g. "Bodyweight", the formula would be easier to read and check. Note the use of embedded constants – they have some clinical meaning and bear removal to a data area where they might be explained. Conversely, and perversely, the constants in the labels (column K) are repeated and not used in column L, which is where the "0.02" in the atropine formula comes from. A complicated undocumented formula for Body Surface Area is unused.

The spreadsheet authors provide the following disclaimers:

> *"The authors have exerted every effort to ensure that the drug dosages set forth are in accordance with current recommendations at the time of publication. The user is urged to check the drug's package insert for any changes in indications and dosages as well as for warnings and precautions. The responsibility is ultimately that of the prescribing clinician".*

We would regard this spreadsheet application as being safety critical and would suggest that there should be independent evidence of the testing to which this spreadsheet has been subjected. The documentation for this spreadsheet comprises one file containing the following information:

> *DIRECTIONS FOR USE OF PEDCAL*

> 1. OPEN FILE.
> 2. SELECT READ ONLY.
> 3. ENTER PATIENT'S NAME IN TOP LEFT HAND CORNER AND ENTER DATA IN "DATA ENTRY BOX." COMPLETE AS MUCH INFORMATION AS POSSIBLE. INFORMATION WILL HIGHLIGHT RED WHEN PROPERLY ENTERED. ENTER AGE IN ONE FIELD ONLY, EITHER MONTHS OR YEARS.
> 4. PRINT.
> 5. GO TO FILE/EXIT. IN SAVE CHANGES DIALOG BOX CLICK "NO."

We would regard this level of documentation as being inadequate in a safety critical software application.

## 3 SOME INITIAL TESTING

We took the opportunity to test a small number of spreadsheets using the "SpACE methodology" [HMRC, 2006]. One was the paediatric anaesthesia model pediatriccalV2.xls introduced above. The other two were taken from several thousand posted on www.medal.org, the website of the Institute for Algorithmic Medicine. One of these is intended to assess the risk of cardiac problems arising in patients undergoing non-cardiac surgery [Cardiac, 2006], and the other to support a decision to assess an elderly patient for masked depression [Svirbely, 2006].

Material errors in any of these could have catastrophic consequences for the patients concerned.

Our testing was confined to spreadsheet use and mechanics. Without the required domain knowledge it is not possible to comment on the appropriateness / completeness of data inputs, the appropriateness of their use, dosage and other interpretive issues. The spreadsheet testing we performed produced over 15 pages of detailed notes, which we omit for clarity.

### 3.1 Findings

Our knowledge of the clinical domain was not sufficient to determine whether any material clinical errors were present. However, we noted a very alarming incidence of poor / high risk practice in the spreadsheet modelling performed.

3.1.1. Common to all three models was extensive use of:

- Constants for drug dosage, risk factor scores, and predicted body measurements embedded in formulas
- Complex nested IF formulas (that shown above was by no means the most complex) some with multiple AND and OR conditions. Many of these also had embedded constants
- Protection / locked / unlocked cells.

- Formulas with no dependents – Some were completeness checks, but some appeared to be potentially important calculations.

3.1.2. Lacking in all three were:

- Documentation of the spreadsheets' workings, and instructions for their use was not present **in the relevant Excel files.** Limited instructions were given in the source web sites, but this was not linked to or embedded in the application.
- The use of data validation to ensure that accurate and appropriate data was input to the models.

3.1.3 Paediatric Anaesthesia model only

- Most dosage information was given in milligrams. A very few doses were shown in micrograms (in the labels) but in a column headed mg (expressed in 3 decimals).

**3.2 Discussion of broad findings**

**3.2.1 Embedded Constants**

These make maintenance very difficult, and hide the internal workings of the spreadsheet from users. Experience in other domains shows that if the constants were to change (perhaps because a drug company changes its recommended dosage, or there is a change in clinical practice) there is a very high risk that the spreadsheet would not be changed to reflect this.

Because the constants are hidden in the formulas, there is no easy way for a user to confirm that the embedded values are the same as those shown in the adjacent labels.

This is important because as discussed above, it is very easy to circumvent the protection and change the formulas, either by accident or by design. Because these spreadsheets are distributed freely over the internet there is no way that all users can be identified for a "product recall" in the event that an error is discovered or an update is found to be necessary.

It would be much more secure if the spreadsheets used formulas that "looked up" external values clearly identified elsewhere on the spreadsheet.

**3.2.2. Complex Nested IF formulas**

These are used to determine dosages or risk factors relating to multiple variables. In the spreadsheets examined, they commonly have multiple conditions and complex conditional logic. This class of formulas is known to be among the most error-prone and difficult to maintain.

A more robust and secure alternative would be to use the VLOOKUP function to determine the value. This would be more transparent, easier to maintain, and present a much lower risk of error.

### 3.2.3. Protection / Locked cells

The spreadsheets examined use protection in conjunction with appropriately locked / unlocked input cells to give some elementary security. As discussed above, this is easy to circumvent and may provide a false sense of security to users, especially given the complex formulas and embedded numbers outlined above.

### 3.2.4 Documentation

Elementary (but, as stated above, inadequate) user instructions and background information was given on the web sites from which the spreadsheets tested were obtained. This was not repeated or linked to in the models.

Human nature will inevitably lead users of these spreadsheets who find them useful to distribute them to colleagues. Without adequate documentation, there is a high risk of inappropriate use. While registered users of some web sites may receive advice of updates and corrections there is a high risk that users who have received them at second hand will not.

### 3.2.5. Data Validation

The two spreadsheets from medal.org included some basic completeness checking, displaying error messages if blank cells are present in the input range. These give no assurance that correct values are present, as they would fail to detect (for example) spaces or text placed in cells where numbers are expected. There is no check to detect nonsense values, or input errors (for example, the dosages calculated in the paediatric anaesthesia model depend on age and body weight, but will not detect such unfeasible input as a 300 lb, 6 ft tall, 6 month baby. This is a gross example – We suspect that much subtler input errors could cause disastrous errors in dosage

The correct use of Excel's Data Validation functions or of its forms tools to restrict the available inputs and warn about unexpected or out-of-range values would greatly reduce the risk of error from incorrect inputs in these critical areas.

### 3.2.6. Formulas with no dependents

All the spreadsheets examined contained formulas with no dependents that did not appear to be the end results. Our domain knowledge is not sufficient to allow us to determine whether these are critical errors or merely ways of displaying optional information that is "nice to have" but is not directly related to a model's purpose.

### 3.2.7 Units in the Paediatric Anaesthesia model

While the units are clearly marked in the labels, it would be fairly easy for a practitioner to confuse milligrams and micrograms, resulting in a serious potential over or under dose of medication, especially where microgram quantities are displayed in a column headed "mg" (see the Clonidine line (row 9) in the spreadsheet extract shown above).

The US Institute of Medicine [Kohn et al (2000)] states that in the US "Medication errors…account for over 7,000 deaths annually" They also cite individual tragic and avoidable deaths that have been caused by dosage mix-ups.

The minimum security against errors that should be introduced is to display drugs with exceptions from the normal dosage units with a different background colour to draw users' attention to the difference.

### 3.3 Spreadsheet Health Risks Identified

Not all of the risks identified arise from the mechanical aspects of spreadsheets. Many of them are potentially compounded by the method of distribution adopted, which allows:

- Distribution of the spreadsheet models separately from the instructions
- Onward distribution of models to individuals not registered with the providing web site. This means that the authors cannot issue updated models or "product recalls" to all users
- Amendment and onward distribution of medical spreadsheets with a spurious "seal of approval" from the originators.

### 4 SUMMARY AND CONCLUSION

We acknowledge the clinical abilities and knowledge of those involved in the development of the spreadsheets we have examined. We applaud their intent to make available the highest standard of medical care, and we regret that we have to identify specific examples. Our purpose is to highlight the following points based upon our own knowledge of the spreadsheet domain:

- The risks arising from use of untested and poorly engineered spreadsheets in clinical medicine and
- The apparent lack of (and therefore the scope for development of) good practice in developing spreadsheet models for clinical use.

Source material for good practice in spreadsheet development has been developed for the financial and taxation domains and is widely and freely available [Read, N. & Batson, J., 1999] [O'Beirne, 2005]. We call for more research into the use of spreadsheets in this safety-critical domain and for the porting of spreadsheet good practice from the financial area to the medical profession.

Documentation relating to the use of spreadsheets in clinical medicine invariably states that ultimate responsibility for the use of such spreadsheets lies with the user. Such is the case in business where the spreadsheet user, often a highly qualified and experienced business professional, bears ultimate responsibility. We believe, in each case, that delegation of responsibility is no barrier to the repeated perpetration of grave errors.